A fast, invariant representation for human action in the visual system

Leyla Isik[*], Andrea Tacchetti[*], and Tomaso Poggio

Center for Brains, Minds, and Machines, MIT

Corresponding author: Leyla Isik

77 Massachusetts Avenue

Bldg 46-4141B

Cambridge, MA 02139

617.258.6933

lisik@mit.edu

Running title: Invariant neural representations for human action

---

[*] These authors contributed equally to this work

**Abstract**

Humans can effortlessly recognize others' actions in the presence of complex transformations, such as changes in viewpoint. Several studies have located the regions in the brain involved in invariant action recognition, however, the underlying neural computations remain poorly understood. We use magnetoencephalography (MEG) decoding and a dataset of well-controlled, naturalistic videos of five actions (run, walk, jump, eat, drink) performed by different actors at different viewpoints to study the computational steps used to recognize actions across complex transformations. In particular, we ask when the brain discounts changes in 3D viewpoint relative to when it initially discriminates between actions. We measure the latency difference between invariant and non-invariant action decoding when subjects view full videos as well as form-depleted and motion-depleted stimuli. Our results show no difference in decoding latency or temporal profile between invariant and non-invariant action recognition in full videos. However, when either form or motion information is removed from the stimulus set, we observe a decrease and delay in invariant action decoding. Our results suggest that the brain recognizes actions and builds invariance to complex transformations at the same time, and that both form and motion information are crucial for fast, invariant action recognition.

**Keywords**

As a social species, humans rely on recognizing the actions of others in their everyday lives. We quickly and effortlessly extract action information from rich dynamic stimuli, despite variations in the visual appearance of action sequences, due to transformations such as changes in size, position, actor, and viewpoint (e.g., is this person running or walking towards me, regardless of which direction they are coming from). The ability to recognize actions, the middle ground between action primitives (e.g., raise the left foot and move it forward) and activities (e.g., playing basketball) (Moeslund and Granum 2001), is paramount to humans' social interactions and even survival. The computations driving this process, however, are poorly understood. This lack of computational understanding is evidenced by the fact that even state of the art computer vision algorithms, convolutional neural networks, which match human performance on object recognition tasks (He et al. 2015), still drastically underperform humans on action recognition tasks (Le et al. 2011; Karpathy et al. 2014). In particular, what makes action and other visual recognition problems challenging are transformations (such as changes in scale, position and 3D viewpoint) that alter the visual appearance of actions, but are orthogonal to the recognition task (DiCarlo and Cox 2007).

Several studies have attempted to locate the regions in the brain involved in processing actions, and in some cases, locate regions in the brain containing viewpoint-invariant representations. In humans and nonhuman primates, the extrastriate body area (EBA) has been implicated in recognizing human form and action (Michels et al. n.d.; Downing et al. 2001; Lingnau and Downing 2015), and the superior temporal sulcus (STS) has been implicated in recognizing biological motion and action (Perrett et al. 1985; Oram and Perrett 1996; Grossman et al. 2000; Vaina et al. 2001; Grossman and Blake 2002; Beauchamp et al. 2003; Peelen and Downing 2005; Vangeneugden et al. 2009). The posterior portion of the STS (pSTS) represents particular types of biological motion data in a viewpoint invariant manner (Grossman et al. 2010; Vangeneugden et al. 2014). Beyond visual cortex, action representations have been found in human parietal and premotor cortex when people perform and view certain actions, particularly hand grasping and goal-directed behavior (analogous to monkey "mirror neuron" system) (Hamilton and Grafton 2006; Dinstein, Gardner, et al. 2008; Dinstein, Thomas, et al. 2008; Oosterhof et al. 2010, 2012, 2013; Freeman et al. 2013). However, recent work suggests that these "mirror neuron" regions do not code the abstract, invariant representations of actions, which are coded in occipitotemporal regions (Wurm et al. 2015, 2016).

Here we investigate the neural dynamics of action processing, rather than the particular brain regions involved, in order to elucidate the underlying computations. We use magnetoencephalography (MEG) decoding to understand when action information is present and how the brain computes representations that are invariant to complex, non-affine transformations such as changes in viewpoint. Timing information can constrain the computations underlying visual recognition by informing when different visual representations are computed. For example, recent successes in MEG decoding have revealed interesting properties about invariant object recognition in humans, mainly that it is fast and highly dynamic, and that varying levels of abstract categorization and invariance increase over the first 200ms following image onset (Carlson et al. 2011, 2013; Cichy et al. 2014; Isik et al. 2014).

Specifically, we use timing data to distinguish between two competing hypotheses of how the brain computes invariance to complex, non-affine transformations. The first hypothesis is that an initial recognition module identifies which action is being performed, and then invariance to various transformations is computed at a later processing stage by discounting specific poses or actor identity information. This hypothesis is consistent with previous data for invariant face and object recognition where shapes can first be discriminated without any generalization and then invariance develops at later stages in the visual processing hierarchy (Logothetis and Sheinberg 1996; Freiwald and Tsao 2010), and is also the basis for viewpoint-invariant computational models of vision (Leibo et al. 2017). Alternatively, the brain may carry out recognition and invariance computations simultaneously, in which case actions could be decoded at the same time within and across different transformations.

Our results show no difference in decoding latency or temporal profile between invariant and non-invariant action recognition, supporting the hypothesis that the brain performs recognition and invariance in the same processing stage. We further show that two types of action information, form (as tested with static images) and motion (as tested with point light figures), both contribute to these immediately view-invariant representations; when either form or motion information is removed, view-invariant decoding is no longer computed at the same time as non-invariant decoding. These results suggest that features that are rich in form and motion content drive the fast, invariant representation of the actions in the human brain.

**Materials and Methods**

*Action recognition dataset*

To study the effect of changes in view on action recognition, we used a dataset of five actors performing five different actions (drink, eat, jump, run and walk) on a treadmill from two different views (0 and 90 degrees from the front of the actor/treadmill; the treadmill rather than the camera was rotated in place to film from different viewpoints) [Figure 1] (Tacchetti et al. 2016). These actions were selected to be highly familiar, and thus something subjects would have experienced under many viewing conditions, to include both reaching-oriented (eat and drink) and leg-oriented (jump, run, walk) actions, as well as to span both coarse (eat and drink versus run and walk) and fine (eat versus drink and run versus walk) action distinctions. Every video was filmed on the same background, and the same objects were present in each video, regardless of action. Each action-actor-view combination was filmed for at least 52-seconds. The videos were then cut into two-second clips that each included at least one cycle of each action, and started at random points in the cycle (for example, a jump may start midair or on the ground). This dataset allows testing of actor and view invariant action recognition, with few low-level confounds.

To explore the roles of form and motion in invariant action representations, we extended this video dataset with two additional components: a form only dataset, consisting of representative single frames for each action, and a motion-only dataset, consisting of point light figures performing the same actions. For the form dataset, the authors selected one frame per video making sure that the selected frames were unambiguous for action identity (special attention was paid to the actions eat and drink to ensure the food or drink was near the mouth, and occluded views to ensure there was some visual information about action). For the motion point light dataset, the videos were put on Amazon Mechanical Turk and workers were asked to label 15 landmarks in every single frame: center of head, shoulders, elbows, hands, torso, hips, knees, and ankles. Three workers labeled each video frame. We used the spatial median of the three independent labels for each frame and landmark to increase the signal to noise ratio, and independently low-pass filtered the time series (Gaussian Filter with a 30 frames aperture and normalized convolution) for each of the 15 points to reduce the high frequency artifacts introduced by single-frame labeling.

*Participants*

Three separate MEG experiments were conducted (see below). Ten subjects (5 female) participated in experiment one, ten subjects (7 female) participated experiment two, and ten subjects (7 female) participated in experiment three. All subjects had normal or corrected to normal vision. The MIT Committee on the Use of Humans as Experimental Subjects approved the experimental protocol. Subjects provided informed written consent before the experiment.

*Experimental procedure*

In the first experiment, we assessed if we could read out different actions both within viewpoint (training and testing on videos at 0 degrees or 90 degrees, without any generalization) and across viewpoint, by training and testing on two different views (0 and 90 degrees). In this experiment ten subjects were shown 50 two-second video clips (one for each of five actors, actions, and two views, 0 and 90 degrees), each presented 20 times.

To examine whether form and motion information were necessary to construct invariant action representations, in the second and third experiments we showed subjects limited "form" (static image) or "motion" (point-light walkers) datasets. Specifically, in the second experiment, ten subjects were shown 50 static images (one for each of five actors, actions, and two views, 0 and 90 degrees), which were single frames from the videos in Experiment 1, for 2 seconds presented 20 times each. In the third experiment, ten subjects were shown 10 two-second video clips, which consisted of point-light walkers traced along one actor's videos from two views in experiment one (labelled by Mechanical Turk workers as described above), presented 100 times each.

In each experiment, subjects performed an action recognition task, where they were asked after a random subset of videos or images (in a randomly interspersed 10% of the trials for each video or image) what action was portrayed in the previous image or video. The purpose of this behavioral task was to ensure subjects were attentive and assess behavioral performance on the various datasets. The button order for each action was randomized across trials to avoid systematic motor confounds in the decoding. Subjects were instructed to fixate centrally. The videos were presented using Psychtoolbox to ensure accurate timing of stimulus onset. Each video had a duration of 2s and a 2s inter-stimulus interval. The videos were shown in grayscale at 3 x 5.4 degrees of visual angle on a projector with a 48 cm × 36 cm display, 140 cm away from the subject.

*MEG data acquisition and preprocessing*

The MEG data were collected using an Elekta Neuromag Triux scanner with 306 sensors, 102 magnetometers at 204 planar gradiometers, and were sampled at 1000 Hz. First the signals were filtered using temporal Signal Space Separation (tSSS) with Elekta Neuromag software. Next, Signal Space Projection (SSP) (Tesche et al. 1995) was applied to correct for movement and sensor contamination. The MEG data were divided into epochs from -500 - 3500 ms, relative to video onset, with the mean baseline activity removed from each epoch. The signals were band-pass filtered from 0.1–100 Hz to remove external and irrelevant biological noise (Acunzo et al. 2012; Rousselet 2012). The convolution between signals and bandpass filter was implemented by wrapping signals in a way that may introduce edge effects at the beginning and end of each trial. We mitigated this issue by using a large epoch window (-500-3500 ms) and testing significance in a manner that takes into account temporal biases in the data (see significance testing below). The above pre-processing steps were all implemented using the Brainstorm software (Tadel et al. 2011).

*General MEG decoding methods*

MEG decoding analyses were performed with the Neural Decoding Toolbox (Meyers 2013), a Matlab package implementing neural population decoding methods. In this decoding procedure, a pattern classifier was trained to associate the patterns of MEG data with the identity of the action in the presented image or video. The stimulus information in the MEG signal was evaluated by testing the accuracy of the classifier on a separate set of test data. This procedure was conducted separately for each subject and multiple re-splits of the data into training and test data were utilized.

The time series data of the magnetic field measured in each sensor (including both the magnetometers and gradiometers) were used as classifier features. We averaged the data in each sensor into 100 ms overlapping bins with a 10 ms step size, and performed decoding independently at each time point. Decoding analysis was performed using cross validation, where the dataset was randomly divided into five cross validation splits. The classifier was then trained on data from four splits (80% of the data), and tested on the fifth, held out split (20% of the data) to assess the classifier's decoding accuracy.

*Decoding - feature pre-processing*

To improve signal to noise, we averaged together the different trials for each semantic class (e.g. videos of run) in a given cross validation split so there was one data point per stimulus per cross validation split. We next Z-score normalized that data by calculating the mean and variance for each sensor using only the training data. We then performed sensor selection by applying a five-way ANOVA to each sensor's training data to test if the sensor was selective for the different actions. We use sensors that were selective for action identity, i.e., show a significantly greater variation across class than within class, with $p < 0.05$ significance based on a F-test (if no sensors were deemed significant, the one with the lowest p-value is selected). The selected sensors were then fixed and used for testing. Each sensor (including both magnetometers and gradiometers) was considered as an independent sensor input into this algorithm, and the feature selection, like the other decoding steps is performed separately at each 100ms time bin, and thus a different number of sensorswas selected for each subject at each time bin. The average number of sensors selected for each subject across all significant decoding time bins is shown in Supplemental Table 2.  These pre-processing parameters have been shown to empirically improve MEG decoding signal to noise in a previous MEG decoding study (Isik et al. 2014), however as we did not use absolute decoding performance (rather significantly above chance decoding) as a metric for when information is present in the MEG signals, we did not further optimize decoding performance with the present data.

*Decoding - classification*

The pre-processed MEG data was then input into the classifier. Decoding analyses were performed using a maximum correlation coefficient classifier, which computed the correlation between each test vector and a mean training vector that is created from taking the mean of the training data from a given class. Each test point was assigned the label of the class of the training data with which it was maximally correlated. When we refer to classifier "training" this could alternatively be thought of as learning to discriminate patterns of electrode activity between the different classes in the training data, rather than a more involved training procedure with a more complex classifier. We intentionally chose a very simple algorithm to see in the simplest terms what information is coded in the MEG data. Prior work has also shown empirically that results with a correlation coefficient classifier are very similar to standard linear classifiers like support vector machines (SVMs) or regularized least squares (RLS) (Isik et al. 2014).

We repeated the above decoding procedure over 50 cross validation splits, at each time bin to assess the decoding accuracy versus time. We measured decoding accuracy as the average percent correct of the test set data across all cross-validation splits, and reported decoding results for the average of ten subjects in each experiment. Plots and latency measures were centered at the median value of each of the 100ms time bins.

For more details on these decoding methods see (Isik et al. 2014).

*Decoding invariant information*

To see if information in the MEG signals could generalize across a given transformation, we trained the classifier on data from subjects viewing the stimuli under one condition (e.g. 0-degree view) and tested the classifier on data from subjects viewing the stimuli under a separate, held out condition (e.g. 90-degree view). This provided a strong test of invariance to a given transformation. In all three experiments, we compared the within and across view decoding. For the "within" view case, the classifier was trained on 80% of data from one view, and tested on the remaining 20% of data from the same view. For the "across" view case, the classifier was trained on 80% of data from one view, and tested on 20% of data from the opposite view, so the same amount of training and test data was evaluated in each case.

*Significance testing*

We assessed action decoding significance using a permutation test. We ran the decoding analysis for each subject with the labels randomly shuffled to create a null distribution. Shuffling the labels breaks the relationship between the experimental conditions that occurred. We repeated the procedure of shuffling the labels and running the decoding analysis 100 times to create a null distribution, and reported p-values as the percentage rank of the actual decoding performance within the null distribution.

For each experiment and decoding condition, we averaged the null decoding data across ten subjects and determined when the mean decoding across subjects was above the mean null distribution. We define the decoding "onset time" as the first time the subject-averaged decoding accuracy was greater than the subject-averaged null distribution, with $p < 0.05$. This provided a measure of when significant decodable information was first present in the MEG signals, and is a standard metric to compare latencies between different conditions (Isik et al. 2013; Cichy et al. 2016). Time of peak decoding accuracy for each condition, an alternative established measure of decoding latency, was found to be much more variable (with 95% CI that were on average over 400 ms larger than onset times), we therefore restricted ourselves to using onset latency only.

To compare onset latencies for different decoding conditions (e.g. within view versus across view decoding), we performed 1000 bootstrap resamples of subjects and use the resulting distribution to compute empirical 95%-confidence intervals for the onset latency of ech condition to estimate the temporal sensitivity of our measure (Hoenig and Heisey 2001), as well as for the difference in onset latency between the two conditions. Specifically, in each bootstrap run, we randomly selected a different subset of ten subjects with replacement, computed onset latencies for each condition (as outlined above) and calculated the difference in onset latency between the invariant and non-invariant conditions. We defined the onset latencies for invariant and non-invariant decoding significantly different with $p<0.05$ if the empirical 95% interval for their difference did not include 0 (Cichy et al. 2016).

*Temporal Cross Training*

Beyond decoding latency, we sought to examine the dynamics of the MEG decoding using temporal-cross-training analysis (Meyers et al. 2008; Meyers 2013; Isik et al. 2014; King and Dehaene 2014). In this analysis, rather than training and testing the classifier on the same time point, a classifier was trained with data from one time point and then tested on data from all other time points. Otherwise the decoding methods (including feature pre-processing, cross validation and classification) were identical to the procedure outlined above. This method yielded a matrix of decoding accuracies for each training and test time point, where the rows of the matrix indicate the times when the classifier was trained, and the columns indicate the times when the classifier was tested. The diagonal entries of this matrix contained the results from when the classifier was trained and tested on data from the same time point (identical to the procedure described above).

**Results**

*Readout of actions from MEG data is early and invariant*

Ten subjects viewed 2-second videos of five actions performed by five actors at two views (0 degrees and 90 degrees) (Figure 1, top row) while their neural activity was recorded in the MEG. We then trained our decoding classifier on only on one view (0 degrees or 90 degrees), and tested it on the second view (0 degrees or 90 degrees). Action can be read out from the subjects' MEG data in the case without any invariance ("within view" condition) at, on average, 250 ms (210-330 ms) (mean decoding onset latency across subjects based on $p < 0.05$ permutation test, 95% CI reported throughout in parentheses, see Methods) [Figure 3, blue trace, Table 1].

The onset latency was much less than the 2s long stimulus and the cycle of most actions (drink, eat, jump which were filmed to each last approximately 2s) in the dataset. Each video began at a random point in a given action sequence, suggesting that the brain can compute this representation from different partial sequences of each action. We also observed a significant rise in decoding after the video offset, consistent with offset responses that have been observed in MEG decoding of static images (Carlson et al. 2011).

We assessed if the MEG signals are invariant to changes in viewpoint by training the classifier on data from subjects viewing actions performed at one view and testing on a second held out view. This invariant "across-view" decoding arose on average at 230 ms (220-270ms). The onset latencies between the within and across were not significantly different at the group level ($p = 0.13$), suggesting that the early action recognition signals are immediately view invariant.

To ensure that the lack of latency difference between the within and between view conditions is not due to the fact that we are using 100ms overlapping time bins, we re-ran the decoding 10ms time bins and 10ms step size (non-overlapping time bins) [Supplemental Figure 3]. Although the overall decoding accuracy is lower, the within and across view decoding onsets were still not significantly different ($p = 0.62$).

*The dynamics of invariant action recognition*

Given that the within- and across-view action decoding conditions ha similar onset latencies, we further compared the temporal profiles of the two conditions by asking if the neural codes for each condition are stable over time. To test this, we trained our classifier with data at one time point, and tested the classifier at all other time points. This yielded a matrix of decoding accuracies for different train times by test times, referred to as a temporal cross training (TCT) matrix (Meyers et al. 2008; Carlson et al. 2013; Meyers 2013; Isik et al. 2014). The diagonal of this matrix shows when the classifier is trained and tested with data at the same time point, just as the line plots in Figure 3.

The within-view and across-view TCTs show that the representations for actions, both with and without view, are highly dynamic as there is little off-diagonal decoding that is significantly above chance (Figure 4a-b, Supplemental Figure 4a-b). The window of significantly above chance decoding performance from 200-400 ms, in particular, is highly dynamic and decoding only within a 50-100 ms window is significantly above chance. At later time points, the above chance decoding extends to a larger window that spans 300ms, suggesting the late representations for action are more stable across time than the early representations. Further, we find that significant decoding for the within and across view conditions are largely overlapping (Supplemental Figure 4C) showing that information for both conditions are represented at the same time scale in the MEG data.

*Coarse and fine action discrimination*

We examined how the five different actions are decoded in both the within and across decoding conditions. By analyzing the confusion matrices for the within- and across-view decoding, we see that not only are coarse action distinctions made (e.g., between run/walk and eat/drink), but so are fine action distinctions (e.g., between eat and drink) even at the earliest decoding of 250 ms (Supplemental Figure 2). Further, actions performed in a familiar context (i.e. run and walk on a treadmill) are not better classified than those performed in an unfamiliar context (i.e. eat and drink on a treadmill).

To more fully examine the difference between coarse and fine action distinctions, we performed decoding on four action pairs - two action pairs that spanned body parts and can be done based on coarse discrimination (drink/run and eat/walk), and two action pairs involving the same body parts and require fine discrimination (run/walk and eat/drink), (Figure 4, Supplemental Figure 5). This decoding is above chance in three of the four pairs (excluding, eat/drink), and as expected, the coarse action distinctions are better decoded than the fine action distinctions. Importantly, in all four pairs there is no significant difference between the latencies for the within and across view decoding (Table 1), strengthening the notion that the representation of action sequence computed by the human brain is immediately invariant to changes in 3D viewpoint.

*The roles of form and motion in invariant action recognition*

To study the roles of two information streams, form and motion, in action recognition, subjects viewed two limited stimulus-sets in the MEG. The first 'Form' stimulus set consisted of one static frame from each video (containing no motion information). The second 'Motion' stimulus set, consisted of point light figures that are comprised of dots on each actor's head, arm joints, torso, and leg joints and move with the actor's joints (containing limited form information) (Johansson 1973).

Ten subjects viewed each of the form and motion datasets in the MEG. We could decode action from both datasets in the within view case without any invariance (Figure 5). The early view-invariant decoding that was observed with full movies, however, was impaired for both the form or motion datasets. In the form-only experiment, within view could be read out at 410 ms and across view at 510ms. The decoding for both the within and across view conditions, however, were not reliably above chance as the lower limit of the 95% CI for both the within and across view decoding fell below the significance threshold. In the motion-only experiment, within view action information could be read out significantly earlier than across view information: 210 ms (180-260 ms) versus 300 ms (300-510 ms), and was significantly different between the two conditions (p = 0.013).

**Discussion**

We investigated the dynamics of invariant action recognition in the human brain and found that action can be decoded from MEG signals around 200 ms post video onset. This is extremely fast, particularly given that the duration of each video and most action cycles (e.g., one drink from a water bottle) was 2s. These results are consistent with a recent MEG decoding study that classified two actions, reaching and grasping, slightly after 200ms post video onset (Tucciarelli et al. 2015). Crucially, we showed that these early neural signals are selective to a variety of full-body actions as well as invariant to changes in 3-D viewpoint.

The fact that invariant and non-invariant action representations are computed at the same time may represent a key difference between the neural computations underlying object and action recognition. Invariant object information increases along subsequent layers of the ventral stream (Logothetis and Sheinberg 1996; Rust and Dicarlo 2010) causing a delay in invariant decoding relative to non-invariant decoding (Isik et al. 2014). Further, physiology data (Freiwald and Tsao 2010) and computational models (Leibo et al. 2017) of static face recognition have shown that invariance to 3D viewpoint, in particular, arises at a later processing stage than initial face recognition. One possible account of this discrepancy is that even non-invariant ("within view") action representations rely on higher-level visual features (that carry some degree of viewpoint invariant information), than those used in basic object representations.

We characterized the dynamics of action representations using temporal cross training and found that the decoding windows for within and across view decoding are largely overlapping (Supplemental Figure 4C), suggesting that the beyond onset latencies, the overall dynamics of decoding are similar for non-invariant and view-invariant action representations. It has been suggested that visual recognition, as studied with static object recognition, has a canonical temporal representation that is demonstrated by highly diagonal TCT matrices (King and Dehaene 2014). Our action results seem to follow this pattern (Figure 4). Representations for action are highly dynamic, but they are more stable over time than previously reported for object decoding (Carlson et al. 2013a; Cichy et al. 2014; Isik et al. 2014).

As shown previously, we find that people can recognize actions with either biological motion or form information removed from the stimulus (Johansson 1973; Schindler and van Gool 2008; Singer and Sheinberg 2010), and that decoding actions within-view is largely intact when form or motion cues are removed. This is likely due to the fact that within-view decoding, unlike the across-view condition, requires little generalization and can thus be performed using low-level cues in the form or motion stimuli. The across-view decoding, on the other hand, requires substantially more generalization and cannot be performed as well, or as quickly with form or motion depleted stimuli. While our datasets are a best attempt to isolate form and motion information, it is important to note that static images contain implied motion and that point light figures contain some form information and have less motion information than full movies. Nevertheless, the low-accuracy and delayed decoding with either limited stimulus set suggests that both form and motion information are necessary to build a robust action representation.

Importantly these invariant action representations cannot be explained by low-level stimulus features, such as motion energy (Tacchetti et al. 2016). While we cannot fully rule out the effects of eye movements or shifts in covert attention, eye movement patterns cannot be accounting for our early MEG decoding accuracy, because we do not observe a significant shift in the eye positions between different actions until after 600 ms post video onset and further the same decoder applied to MEG signals does not successfully decode action information using raw eye position data (Supplemental Figure 1).

The five actions tested in this study comprise only a small subset of the wide variety of familiar actions we recognize in our daily lives. The five-way classification shows similar decoding across between all five actions, including both coarse and fine action distinctions (Supplemental figure 2). The binary action decoding results show that, while coarse action discriminations are more easily decoded than fine action discriminations, there is no difference in the latency of within and across view decoding for any action pair, suggesting that invariant and non-invariant action signals are indeed computed at the same stage (Figure 4, Supplemental Figure 5).

These five actions were selected to be highly familiar, and thus we do not know to what extent familiarity is necessary for the immediate invariance we observed. Indeed, modeling and theoretical work suggest that in order to build templates to be invariant to non-affine transformations such as changes in 3-D viewpoint, one must learn templates from different views of each given category (Leibo et al. 2015). It remains an open question how this invariance would translate to unfamiliar actions and how many examples would be needed to learn invariant representations of new actions.

Finally, the longer latency and greater cross-temporal stability of action decoding raises the question of whether recurrent and feedback connections are used to form invariant action representations. This is difficult to test explicitly without high spatiotemporal resolution data.

Beyond 100ms post-stimulus onset it is likely that feedback and recurrent connections occur (Lamme and Roelfsema 2000), however further studies have shown that this data is well approximated by feedforward hierarchical models (Tacchetti et al. 2016).

Taken as a whole, our results show that the brain computes action selective representations remarkably quickly and, unlike in the recognition of static faces and objects, at the same time that it computes invariance to complex transformations that are orthogonal to the recognition task. This contrast represents a qualitative difference between action and object visual processing. Moreover, our findings suggest that both form and motion information are necessary to construct these fast invariant representations of human action sequences. The methods and results presented here provide a framework to study the dynamic neural representations evoked by natural videos, and open the door to probing neural representations for higher level visual and social information conveyed by video stimuli.

## Acknowledgements

This material is based upon work supported by the Center for Brains, Minds and Machines (CBMM), funded by NSF STC award CCF-1231216.  We thank the McGovern Institute for Brain Research at MIT and the Athinoula A. Martinos Imaging Center at the McGovern Institute for Brain Research at MIT for supporting this research. We would like to thank Patrick Winston, Gabriel Kreiman, Martin Giese, Charles Jennings, Heuihan Jhuang, and Cheson Tan for their feedback on this work.

**Figures**

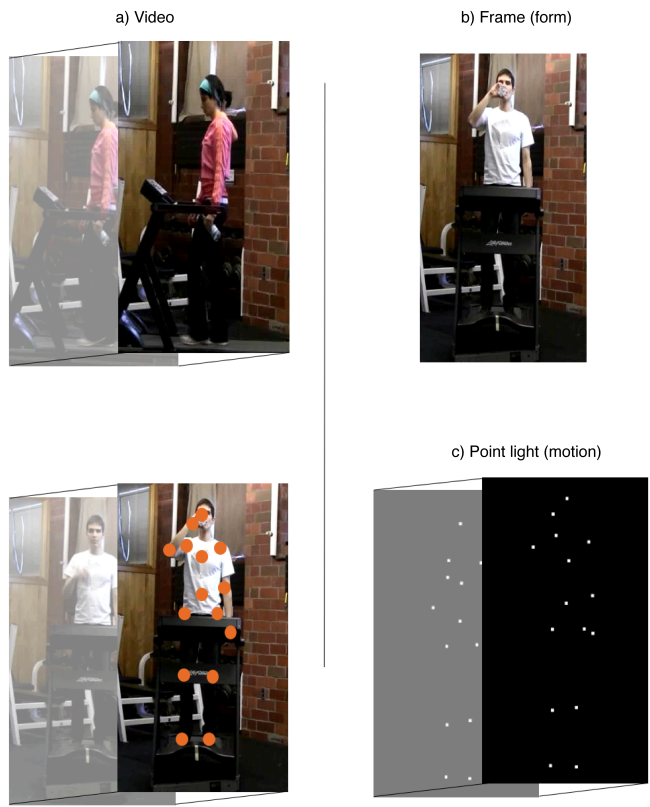

*Figure 1 – Action recognition dataset. A) We used a dataset of two-second videos depicting five actors performing five actions from five viewpoints. Frames from one example walk video at 90 degrees (top) and one example drink video at 0 degrees (bottom) are shown. We extended this dataset to B) a "Form only" dataset, containing single (action informative) frames from each two-second movie, and C) a "Motion only" dataset of point light videos created by labeling joints on actors in each video (A, bottom).*

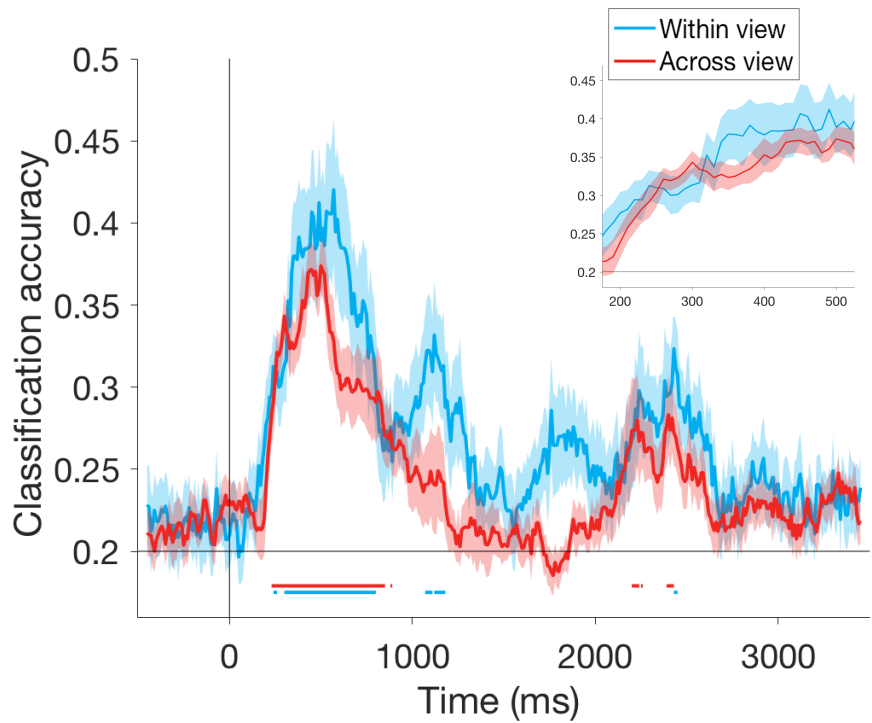

***Figure 2 – Within and across view action decoding.*** *We can decode action by training and testing a simple on the same view ('within-view' condition), or, to assess viewpoint invariance, training on one view (0 degrees or 90 degrees) and testing on second view ('across view' condition). Results are from the average of ten subjects. Error bars represent standard error across subjects. Horizontal line indicates chance decoding accuracy (see Supplementary Materials). Line at bottom of plot indicates group-level significance with p<0.05 permutation test, for the average null distribution across the ten subjects. Inset shows a zoom of decoding time courses from 175-525 ms post-video onset.*

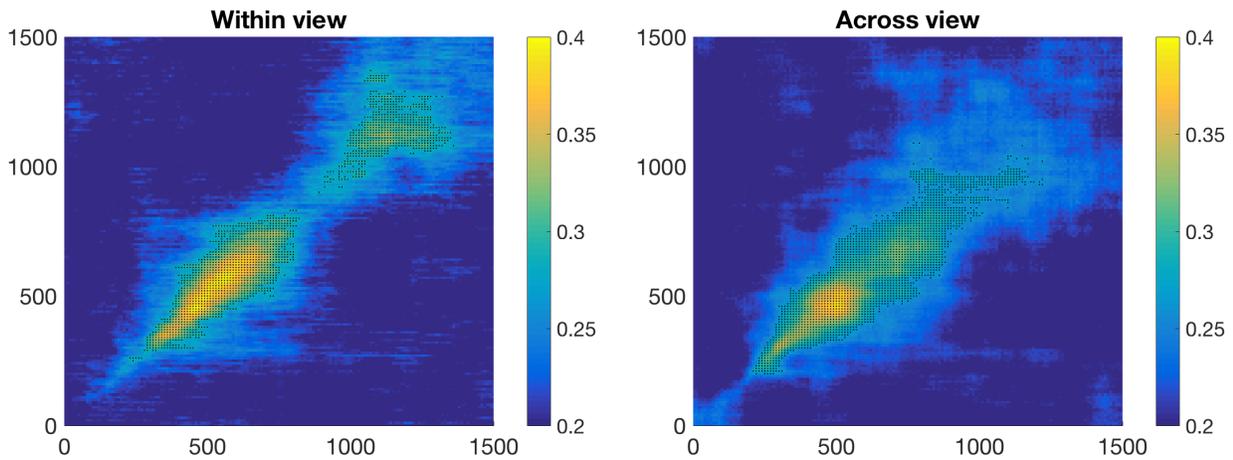

***Figure 3 – Dynamics of action representations.*** *A temporal cross training matrix showing the decoding results for training a classifier at each point in time (y-axis) and testing the classifier at all other times (x-axis), zoomed in to the time period from 0-1500ms post-video onset, for **(a)** within-view decoding, and **(b)** across-view decoding for subjects watching the 2-view video dataset (Experiment 1). Colorbar indicates mean decoding accuracy for ten subjects. Black dots indicate points when decoding is significantly above chance at group level based on p<0.05 significance test. Results along the diagonal for the within and across view decoding are the same as shown in the line plots in Figure 3. For TCT results from -500-3500 ms post-video onset, see Supplemental Figure 4.*

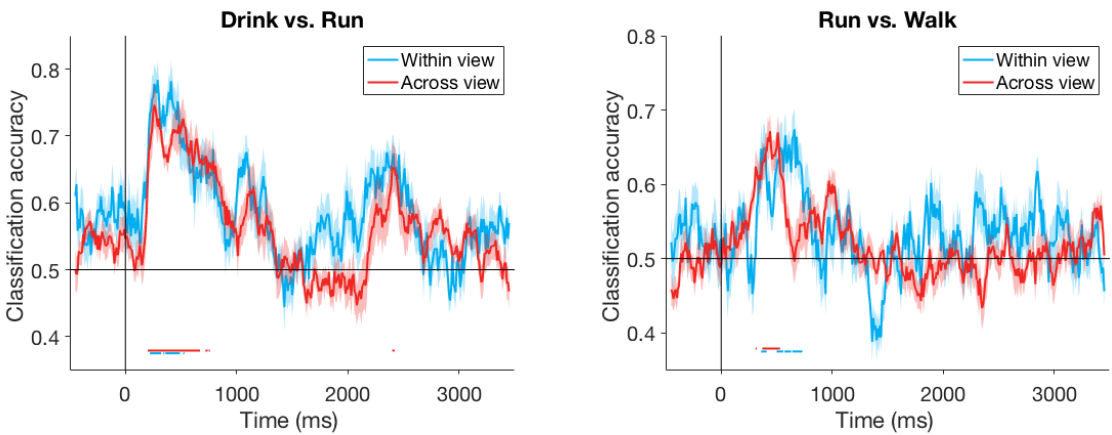

***Figure 4 – Coarse and fine binary action decoding.*** *We can decode action in a pair of dissimilar actions (across body part, coarse discrimination) or a pair of similar actions (within body part, fine discrimination) both 'within-view' and 'across-view'. A) decoding Drink versus Run (coarse discrimination) and b) Run versus Walk (fine discrimination) are shown. Results are from the average of ten subjects watching the 2-view videos dataset (Experiment 1). Error bars represent standard error across subjects. Horizontal line indicates chance decoding accuracy (see Supplementary Materials). Line at bottom of plot indicates group-level significance with p<0.05 permutation test, for the average null distribution across the ten subjects.*

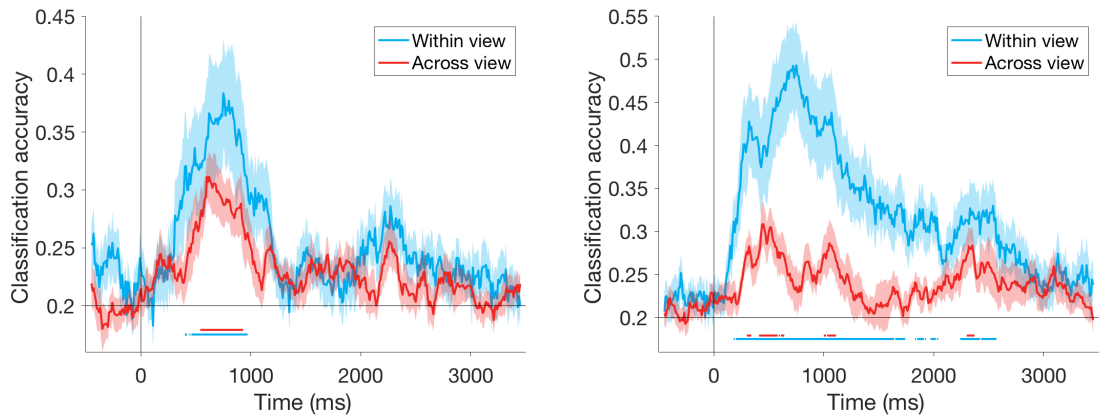

***Figure 5 – The effects of form and motion on invariant action recognition. (a)*** *Action can also be decoded invariantly to view from form information alone (static images)* ***(b)*** *Action can be decoded from biological motion only (point light walker stimuli). Results are each from the average of ten subjects. Error bars represent standard error across subjects. Horizontal line indicates chance decoding (20%). Line at bottom of plot indicates group-level significance with p<0.05 permutation test, for the average null distribution across the ten subjects.*

| Experiment | Within view onset time mean (95% CI) | Across-view onset time mean (95% CI) |
|---|---|---|
| **Experiment 1 (videos) – 100 ms** | 250 ms (210-330) | 230 ms (220-270) |
| **Experiment 1 – 10 ms** | 210 ms (190 – 300) | 250 ms (230-300) |
| **Experiment 1 – Run/Drink** | 230 ms (200-240) | 210 ms (200-260) |
| **Experiment 1 – Run/Eat** | 230 (200-240) | 240 ms (220-270) |
| **Experiment 1 – Run/Walk** | 370 ms (330-430) | 320 ms (280-430) |
| **Experiment 1 – Eat/Drink** | N/A | N/A |
| **Experiment 2 (frames)** | 410 ms (320 --) | 510 ms (430--) |
| **Experiment 3 (point lights)** | 210 ms (180-260) | 300 ms (290-420) |

*Table 1 – Onset times and 95% CI for each experiment*. The onset time for the average across all 10 subjects' data for each experiment, and the 95% CI for each onset time as calculated with a bootstrapping procedure (see Methods).